\documentstyle[preprint,aps]{revtex}

\begin{document}
\preprint{
\font\fortssbx=cmssbx10 scaled \magstep2
\hbox to \hsize{
\hfill$\raise .5cm\vtop{
\hbox{hep-ph/9706539}
}$}
}
\draft
\vfill
\title{Weak Decay Process of $B \to \rho\ \ell\ \bar\nu_\ell$:\\
A Varying External Field Approach in QCD Sum Rules}
\draft
\vfill
\author{Kwei-Chou Yang}
\address{Institute of Physics, Academia Sinica, Taipei, Taiwan 115, R.O.C.}
\maketitle
\begin{abstract}
 A varying external field approach in QCD sum rules is formulated
in a systematic way to treat the weak decay form factors and their
$q^2$ dependence in the process of $B \to \rho\ \ell\ \bar\nu_\ell$.
From the form factor sum rules, we can also obtain the mass sum rules
for $B$ and $\rho$ mesons, which can help us determine the
reliable Borel windows in studying the relevant form factor sum rules. In this
way, we thus demonstrate that some QCD sum rule calculations in the
literature are less reliable. We also include induced condensate
contributions, which have been ignored, into the relevant sum rules.
We obtain the ratios $\Gamma(\bar B^0\to\rho^+ e^- \bar\nu_e)$/
$\Gamma(\bar B^0\to\pi^+ e^- \bar\nu_e)$$\approx$ 0.94 and
$\Gamma(\bar B^0\to\rho^+ \tau^- \bar\nu_\tau)$/
$\Gamma(\bar B^0\to\pi^+ \tau^- \bar\nu_\tau)$ $\approx$ 1.15.
We apply this approach to re-examine the case of the $D$ meson decay.
\end{abstract}
\section  {INTRODUCTION}
 Semileptonic decays, because of their simplicity, provide an excellent
laboratory where physicists can study the effect of nonperturbative QCD
interactions on the weak decay process. A detailed understanding of these
processes is also essential for determining the magnitudes of CKM
quark-mixing matrix elements.

The weak decay form factors of $B\to \rho\ \ell\ \bar\nu_\ell$ have been
calculated by using various approaches. However, the results obtained from the
traditional sum rules\cite{bbd} by considering a three-point correlation
function with  suitable interpolating current seem to conflict seriously with
others' theoretic results, such as light-cone sum rules\cite{bel,ali},
lattice simulations\cite{lat}, quark models\cite{korner,wsb,isgur1,scora,cheng},
and the external field approach of QCD sum rules\cite{yang}. Recently,
Ball and Braun \cite{ball} re-examined such a process by studying
the light-cone sum rules with the modified $\rho$ meson wave functions.
In their study, all of the soft (non-perturbative) parts are absorbed into
the $\rho$ meson wave functions. The accuracy of their results is dependent
on the shape of the wave functions. Similarly, within the pQCD
approach\cite{yu}, the light meson is described by a phenomenological model
function which can be taken, e.g., from QCD sum rules.

In this work, we shall use the varying external field approach of QCD sum
rules which has been developed earlier\cite{yang} to re-examine the weak decay
form factors for $B \to$ $\rho$ $\ell\ \bar\nu_\ell$ with a complete calculation. This
approach, in spirit, is similar to Ref.\cite{rad}. We will present the idea
of the induced condensates and show how the induced condensates enter the sum
rules.  In the calculation we find an additional contribution from the induced
condensates, which was ignored in Ref.\cite{yang}. Consequently, the
$A_1$ form factor will obtain an additional 10$\%$ contribution from the
induced condensate, while the other form factors will not.
In addition, we also propose a reliable method that can be used to
determine the Borel windows in studying the relevant form factor sum rules.

The rest of this paper is organized as follows. The concept of the induced
nonlocal condensates will be formulated in the part (a) of Sec. II.
In the part (b) of Sec. III, the external field approach of QCD sum rules
will be set up to investigate the various decay form factors of
$B\to\rho\ \ell\ \bar\nu_\ell$. In Sec. IV numerical results and discussion
are presented. Sec.V contains a brief summary.

\section{ THE METHOD}
\subsection{Quark propagator in the presence of an external variable field}

The quark propagator in the external vector (axial vector) field is described
by the additional term, $\Delta {\cal L}^{1(2)}$, in the Lagrangian.
\begin{eqnarray}
\Delta {\cal L}^1(x)=-{\cal V}^\alpha e^{iqx}
J^V_\alpha(x),
\end{eqnarray}
or
\begin{eqnarray}
\Delta {\cal L}^2(x)=-{\cal A}^\alpha e^{iqx}
J^A_\alpha(x),
\end{eqnarray}
where
\begin{eqnarray}
J^V_\alpha=\bar u\gamma_\alpha b,\ \ \ \ \  J^A_\alpha=\bar u\gamma_\alpha
\gamma_5 b.
\end{eqnarray}
Here ${\cal V^\alpha}$ (${\cal A^\alpha}$) and $q^\alpha$ are the amplitude and
momentum of the external vector (axial-vector) field, respectively, $b$ stands
for the b quark field operator, and $u$ is for the u quark field operator.
Hence the quark propagator depicted in Fig. 1(a) can be written down as
\begin{eqnarray}
\langle{\rm T} u^{b}_{\alpha}(x) \bar b^a_\beta(0)\rangle_{\cal V(A) }
=\langle{\rm T} u^{b}_{\alpha}(x) \bar b^a_\beta(0)\rangle ^{pert}_{\cal V(A)}
+\langle: u^{b}_{\alpha}(x) \bar b^a_\beta(0):\rangle^{induced}_{\cal V(A)},
\label{a}
\end{eqnarray}
with the notation $\langle\dots\rangle\equiv\langle0\vert\dots\vert0\rangle$.
Here $a$ and $b$ are the color indices while $\alpha$ and $\beta$ the Lorentz
indices.  The first term depicted in Fig. 1(b), on the right hand side of
Eq. (4), can be calculated from perturbation theory, while the second
term depicted in Fig. 1(c) is the induced condensate defined through this
paper.  Neglecting the radiative corrections, we obtain
\begin{eqnarray}
\langle{\rm T} u^{b}_{\alpha}(x) \bar b^a_\beta(0)\rangle ^{pert}_{\cal V}
=i\delta^{ab} \int {{\rm d^4}p\over(2\pi)^4}e^{-ipx}
{[(\not p +m_u)\not {\cal V}((\not p+\not q)+m_b)]_{\alpha\beta}
 \over (p^2-m^2_u)[(p+q)^2-m_b^2]}.
\end{eqnarray}
Using the identity\cite{yang}:
\begin{eqnarray}
&&12\bar b_\beta^a u_{\alpha}^b\nonumber\\
&&=\Bigl[
1(\bar bu)+\gamma_5(\bar b\gamma_5 u)
+\gamma^\rho(\bar b\gamma_\rho u)
-\gamma^\rho\gamma_5(\bar b\gamma_\rho\gamma_5 u)
+{1\over 2}\sigma^{\rho\tau}(\bar b\sigma_{\rho\tau}u)
\Bigr]_{\alpha\beta}\delta^{ab}\nonumber\\
&&=\Bigr[
1(\bar bu)+\gamma_5(\bar b\gamma_5 u)
+\gamma^\rho(\bar b\gamma_\rho u)
-\gamma^\rho\gamma_5(\bar b\gamma_\rho\gamma_5 u)
+{1\over 2}\sigma^{\rho\tau}\gamma_5(\bar b\sigma_{\rho\tau}\gamma_5 u)
\Bigr]_{\alpha\beta}\delta^{ab},\label{b}
\end{eqnarray}
we now project out Eq. (5) on the Dirac matrix basis ${\bf 1}$, $\gamma_5$,
$\gamma_\rho$, $\gamma_\rho \gamma_5$, $\sigma_{\rho\tau}$.
Thus the induced condensate (the second term on the right hand side of
Eq.~(\ref{a})), with the aid of Eq.~(\ref{b}), is
\begin{eqnarray}
\langle: u^{b}_{\alpha}(x) \bar b^a_\beta(0):\rangle ^{induced}_{\cal V}
= &-&{\delta^{ab}\over 12}\left(
\langle:\bar b(0) u(x):\rangle \delta_{\alpha\beta}+
\langle:\bar b(0)\gamma_5 u(x):\rangle (\gamma_5)_{\alpha\beta}
\right.
\nonumber\\
  &+& \langle:\bar b(0)\gamma_\rho u(x):\rangle
(\gamma^\rho)_{\alpha\beta}-\langle:\bar b(0)\gamma_{\rho}\gamma_5 u(x):\rangle
(\gamma^\rho \gamma_5)_{\alpha\beta}\nonumber\\
&+& \left. {1\over 2}\langle:\bar b(0)\sigma_{\rho\tau} u(x):\rangle
(\sigma^{\rho\tau})_{\alpha\beta}\right)^{induced}_{\cal V},\label{d}
\end{eqnarray}
or
\begin{eqnarray}
\langle: u^{b}_{\alpha}(x) \bar b^a_\beta(0):\rangle ^{induced}_{\cal A}
= &-&{\delta^{ab}\over 12}\left(
\langle:\bar b(0) u(x):\rangle \delta_{\alpha\beta}+
\langle:\bar b(0)\gamma_5 u(x):\rangle (\gamma_5)_{\alpha\beta}
\right.  \nonumber\\
  &+&\langle:\bar b(0)\gamma_\rho u(x):\rangle (\gamma^\rho)_{\alpha\beta}
-\langle:\bar b(0)\gamma_{\rho}\gamma_5 u(x):\rangle
(\gamma^\rho \gamma_5)_{\alpha\beta}\nonumber\\
&+&\left. {1\over 2}\langle:\bar b(0)\sigma_{\rho\tau}\gamma_5 u(x):\rangle
(\sigma^{\rho\tau}\gamma_5)_{\alpha\beta}\right)^{induced}_{\cal A},\label{da}
\end{eqnarray}
where
\begin{eqnarray}
\langle:\bar b(0)\Gamma u(x):\rangle^{induced}_{\cal V(A)}
=\langle{\rm T}\bar b(0)\Gamma u(x)\rangle_{\cal V(A)}
-\langle{\rm T}\bar b(0)\Gamma u(x)\rangle^{pert}_{\cal V(A)}
\end{eqnarray}
with
\begin{eqnarray}
\langle{\rm T}\bar b(0) \Gamma u(x)\rangle _{\cal V}&=&-i{\cal V}^\mu \int
{\rm d}^4 z e^{iqz}\langle {\rm T}(\bar u(z)\gamma_\mu b(z),
\bar b(0)\Gamma u(x))\rangle ,\nonumber\\
\langle{\rm T}\bar b(0) \Gamma u(x)\rangle _{\cal A}&=&-i{\cal A}^\mu \int
{\rm d}^4 z e^{iqz}\langle {\rm T}(\bar u(z)\gamma_\mu \gamma_5 b(z),
\bar b(0)\Gamma u(x))\rangle,
\end{eqnarray}
and
\begin{eqnarray}
\langle {\rm T}\bar b(0)\Gamma u(x)\rangle^{pert}_{\cal V}
=-iN_c \int {{\rm d^4}p\over(2\pi)^4}e^{-ipx}
{{\rm Tr}\left[ (\not p +m_u)\not {\cal V}(\not p+\not q+m_b)\Gamma\right]
 \over (p^2-m^2_u)[(p+q)^2-m_b^2]},
\end{eqnarray}
\begin{eqnarray}
\langle{\rm T} \bar b(0)\Gamma u(x) \rangle ^{pert}_{\cal A}
=-iN_c \int {{\rm d^4}p\over(2\pi)^4}e^{-ipx}
{{\rm Tr}[(\not p +m_u)\not {\cal A}\gamma_5((\not p+\not q)+m_b)\Gamma]
 \over (p^2-m_u^2)[(p+q)^2-m_b^2]}\label{ca}.
\end{eqnarray}
Here $\Gamma$ is the generic notation for the Dirac matrix basis ${\bf 1}$,
$\gamma_5$,
$\gamma_\rho$, $\gamma_\rho \gamma_5$, $\sigma_{\rho\tau}$ and the path-ordered
gauge factor is implied by
\begin{eqnarray}
{\cal G}(0,x)= P{\rm exp} [-ig_s\int^1_0 d \alpha x^\mu A_\mu(\alpha x)].
\end{eqnarray}
In the case of the fixed-point gauge (the Fock-Schwinger gauge)
$x^\mu A_\mu(x)=0$, this factor is equal to unity.

\subsection{The Derivation of QCD Sum Rules}
For the form factors of $B$ $\to$ $\rho$ $\ell$ $\bar\nu_\ell$,
we consider the following two point
Green's function in the external variable vector field $\cal V$
or axial-vector field $\cal A$:
\begin{eqnarray}
\Pi_{\mu\nu}^{\cal V(A)}(p,p';q^2)=i \int d^4x \ e^{ip'x}
\langle T\{j_\mu(x), j_\nu^5(0) \} \rangle_{\cal V(A)},
\end{eqnarray}
where $\langle\dots\rangle\equiv\langle0\vert\dots\vert0\rangle$,
$j_\mu=\bar d\gamma_\mu u$ and
$j_\nu^5=\bar b\gamma_\nu\gamma_5 d$.
The interaction with the external vector field is described by the
additional term, $\Delta \cal L$, in the Lagrangian as shown in Eqs. (1) and (2).
In the following calculation we consider only the amplitude of
$\Pi_{\mu\nu}^{\cal V(A)}$ linear in the external field $\cal V(A)$.
Generally speaking, the coefficients of tensor structures $\Pi_i$ in
$T_{\mu\nu}$ can be expressed as a double dispersion relation form:
\begin{eqnarray}
\Pi_i(p,p';q^2)
=\int ds \int ds'{\rho_i(p,p';q^2)\over (s-p^2)(s'-p'^2)}
+\hbox{subtraction\ terms},
\end{eqnarray}
where the subtraction terms have the form
\begin{eqnarray}
\hbox{subtraction\ terms}
=P_{1\mu\nu}(p^2)\!\int\!{\Delta(s')ds'\over s'-p'^2}+
P_{2\mu\nu}(p'^2)\!\int\!{\Delta(s)ds\over s-p^2}
+P_{3\mu\nu}(p^2,p'^2).
\end{eqnarray}
One finds that such a contribution from the subtraction terms is of
no importance since, after performing the double Borel transform, all of
the subtraction terms, $P_{1\mu\nu}, P_{2\mu\nu}$, and $P_{3\mu\nu}$,
disappear. (For further discussions, see Ref. \cite{smilga})
The double Borel transform $B[\Pi_i]$ on $\Pi_i$\cite{yang} in both
variables $p^2$ and $p'^2$ gives
\begin{eqnarray}
B[\Pi_i]=\int_0^\infty ds \int_0^\infty ds'\ e^{-(s'/
M'^2+s/ M^2)}\ \ \rho_i
\end{eqnarray}
The hadronic side of Eq. (14) is represented as a sum over the hadronic states.
If the Borel masses are chosen properly, the hadronic
side of Eq. (14) will be dominated by the lowest hadronic states with the
contributions from the higher states and the continuum suppressed.
At the hadron level (the right hand side), the
spectral function of $\Pi_{\mu\nu}^{\cal A(V)}$ can be written as
\begin{eqnarray}
\rho_{\mu\nu}^{\cal A}=&&{\cal A}^\alpha{{f_B m_\rho^2}\over
g_\rho} [fg_{\alpha\mu}p_\nu+a_+p_\mu(p+p')_\alpha p_\nu+a_-p_\mu q_\alpha
p_\nu \nonumber\\
&&+\hbox{other terms}\dots]\times\delta(s-m_B^2)\delta(s'-m_\rho^2)+
\hbox{higher states,}\nonumber\\
\rho_{\mu\nu}^{\cal V}=&&{\cal V}^\alpha{{f_B m_\rho^2}\over
g_\rho}[2ig\epsilon_{\alpha\mu\beta\rho}p_\nu p'^\beta p^\rho\nonumber\\
&&+\hbox{other terms}\dots]\times\delta(s-m_B^2)\delta(s'-m_\rho^2)+
\hbox{higher states,}
\end{eqnarray}
where the higher states start from a higher enough value, $s_0^B$
or $s_0^\rho$. The contribution from the higher states could be approximated by
the perturbative part, (which
starts from the value, $ s_0^B$ or $s_0^\rho$) in the OPE of the quark level.
The $``$other terms$"$ in Eq. (18) are the irrelevant tensor structures that are
not taken into account here and thus are suppressed. Here we have adopted the
definitions
\begin{eqnarray}
\langle 0|j_\mu|\rho(p',\lambda)\rangle=&&i{m_{\rho}^2\over
g_{\rho}}\epsilon_{(\lambda)\mu},\nonumber\\
\langle \rho(p',\lambda)|J_\alpha^V-J_\alpha^A|B(p)\rangle
=&&2ig\epsilon_{\alpha\mu\beta\rho}p'^\mu p^\beta
\epsilon_{(\lambda)}^{*\rho}\nonumber\\
&&-f\epsilon_{(\lambda)\alpha}^* -
a_+(\epsilon_{(\lambda)}^*\cdot p)(p+p')_\alpha -
a_-(\epsilon_{(\lambda)}^*\cdot p)q_\alpha,
\end{eqnarray}
with
\begin{eqnarray}
f=&&(m_{\rho}+m_B)A_1,\nonumber\\
 a_+=&&-{A_2\over m_{\rho}+m_B},\nonumber\\
 a_-=&&2m_\rho A,\nonumber\\
A(q^2)=&&{A_0(q^2)-A_3(q^2)\over q^2},\
A_3(q^2)={A_1(q^2)(m_\rho+m_B)\over 2m_\rho}-
{A_2(q^2)(m_B-m_\rho)\over 2m_\rho},\nonumber\\
g=&&{V\over m_\rho+m_B}\nonumber
\end{eqnarray}
in the BSW parametrization\cite{wsb}.

  At the quark-gluon level, the Eq. (14) can be alternatively written as
\begin{eqnarray}
\Pi_{\mu\nu}^{\cal V(A)}(p,p';q^2)=-i \int d^4x \ e^{ip'x}
\langle {\rm Tr}\{ S^{ab}_d (0,x)\gamma_\mu
 S_{ub}^{ba\cal V(A)}(x,0)\gamma_\nu \gamma_5 \}\rangle,
\end{eqnarray}
where \cite{yang,yang2}
\begin{eqnarray}
S^{ab}_{d(ij)}(0,x)=&&\int {d^4p\over(2\pi)^4}e^{ipx}
\left[
{i\delta^{ab}\over \not p-m_d} \right.
+ {i\over 4}{\lambda^n_{ab}\over 2}g_s G_{\mu\nu}^n(0)
{\sigma^{\mu\nu}(\not p + m_d) + (\not p + m_d)
\sigma^{\mu\nu}\over (p^2-m_d^2)^2}\nonumber\\
&&\left. -{iG_{\mu\nu}^n(0)\lambda^n_{ab}\over 4}g_s x^\nu
({1\over \not p-m_d}\gamma^\mu {1\over \not p-m_d})
\right]_{ij}\nonumber\\
&&+:d^a_i(0) \bar d^b_j(0):+ x_\mu:d^a_i(0)(D^\mu\bar d^b_j(0)):+
{x_\mu x_\nu\over 2!}:d^a_i(0)(D^\mu D^\nu \bar d^b_j(0)):\nonumber\\
&&+{x_\mu x_\nu x_\lambda\over 3!}:d^a_i(0)(D^\mu D^\nu D^\lambda \bar d^b_j(0)):
\end{eqnarray}
and $S_{ub}^{ba \cal V(A)}(x,0)$
$\equiv \langle T u^b(x) \bar b^a(0) \rangle _{\cal V (\cal A)}$
is shown in Eq. (4) but also in the background gluon field.
Here we have used the fixed-point gauge, $x^\mu A_\mu^a(x)=0$
\cite{yang2,yang}  for the gluon field and 
$D^\alpha=\partial^\alpha+ig_s {\lambda^\alpha\over 2} A^\alpha_a$.

The Feynman integral for the bare loop can be written as
\begin{eqnarray}
&&\Pi_{\mu\nu}^{\cal V}(p,p';q^2)
=3i{\cal V}^\alpha \int{d^4k\over (2\pi)^4}
  {{\rm T}r[(\not p'+\not k +m_u)\gamma_\alpha
(\not p+\not k+m_b)\gamma_\nu \gamma_5 (\not k +m_d)\gamma_\mu]
\over k^2(k+p')^2[(k+p)^2-m_b^2]},\nonumber\\
&&\Pi_{\mu\nu}^{\cal A}(p,p';q^2)
=3i{\cal A}^\alpha \int{d^4k\over (2\pi)^4}
  {{\rm T}r[(\not p'+\not k+m_u)\gamma_\alpha\gamma_5
(\not p+\not k +m_b)\gamma_\nu\gamma_5(\not k +m_d)\gamma_\mu]
\over k^2(k+p')^2[(k+p)^2-m_b^2]},
\end{eqnarray}
where $m_u^2$ and $m_d^2$ have been neglected. The relevant diagram is shown in
Fig. 2(a). Using the Cutkosky's rule\cite{cut}, the integral can be solved
easily by taking the imaginary part of the quark propagators:
$1/(p^2-m_b^2 +i\epsilon) \rightarrow -2\pi i\delta(p^2-m_b^2)$.
Thus this integral may be recast as a double dispersion relation
\begin{eqnarray}
\Pi_{\mu\nu}^{\cal V(A)}(p,p';q^2)
=-{{\cal V(A)}^\alpha\over 4\pi^2}\int\!\!\!\int_{\Omega}dsds'
{\rho_{\mu\nu\alpha}^{\cal V(A)}
(s,s';q^2)\over (s-p^2)(s'-p'^2)},
\end{eqnarray}
where
\begin{eqnarray}
\rho_{\mu\nu\alpha}^{\cal V}=&&3i\int {d^4k\over {(2\pi)^4}}
(-2\pi i)^3 \delta(k^2) \delta[(k+p')^2] \delta[(k+p)^2-m_b^2)]\nonumber\\
&&\times{\rm T}r[(\not p'+\not k +m_u)\gamma_\alpha
(\not p+\not k+m_b)\gamma_\nu \gamma_5 (\not k +m_d)\gamma_\mu],\nonumber\\
\rho_{\mu\nu\alpha}^{\cal A}=&&3i\int {d^4k\over {(2\pi)^4}}
(-2\pi i)^3 \delta(k^2) \delta[(k+p')^2] \delta[(k+p)^2-m_b^2)]\nonumber\\
&&\times{\rm T}r[(\not p'+\not k+m_u)\gamma_\alpha\gamma_5
(\not p+\not k +m_b)\gamma_\nu\gamma_5(\not k +m_d)\gamma_\mu],
\end{eqnarray}
and the corresponding integration region $\Omega$, which can be solved
via the Landau equation, is specified by $s'>0$ and
$s>m_b^2+m_b^2s'/(m_b^2-q^2)$. Note that if ($m_b^2-q^2)/q^2< s'$,
there may be additional contributions to the above integral because of pinching
of the singularities on non-physical sheets\cite{bbd}. However,
at the hadron level, the contribution from the higher states
is approximated by the perturbative part, which starts from the thresholds
$s_0^B$ and $s_0^\rho$, we therefore study the $q^2$ behaviors of the form factors
in the region: $0\le q^2 $ and ($m_b^2-q^2)/q^2>s_0^\rho$,
where $s_0^\rho$ is the threshold of the excited states, $s'<s_0^\rho$.
In this region it is not necessary to worry about
possible contributions from non-physical sheets.
Considering only the leading power corrections of OPE of the correlation
function Eq. (14), we obtain\cite{ex}
\begin{eqnarray}
\Pi_i(p,p';q^2)
&&=\int_0^{s_0^B}ds\int_0^{s_0^\rho}ds'
{\rho^{pert}_i(p,p';q^2)\over (s-p^2)(s'-p'^2)}+\Pi^{induced}_i
+d_3^i \langle\bar dd\rangle+d_5^i \langle g_s\bar d\sigma Gd\rangle\nonumber\\
&& +d_6^{(1)i} g_s^2\langle\bar dd\rangle^2
+d_6^{(2)i} g_s^2\langle\bar dd\rangle\langle\bar bb\rangle
+d_6^{(3)i} g_s^2\langle\bar dd\rangle\langle\bar uu\rangle +\cdots,
\end{eqnarray}
where
\begin{eqnarray}
\rho_{f}^{pert}=&&{{3s'}\over{4\pi^2\lambda^{5/2}}} \Big(
2s'(2s-2m_b^2-u)\Delta-
\lambda[(m_b^2-q^2)^2-s'q^2]\cr
&&-\lambda  m_b (m_u-m_d)(2s'+2m_b^2-u) \Big),
\end{eqnarray}
\begin{eqnarray}
\rho_{a_\pm}^{pert}=&&{{3s'}\over{4\pi^2\lambda^{7/2}}} \Big(
20\Delta s'[\mp
u(s\pm s'-m_b^2)-2s'm_b^2]\cr
&&\mp\lambda[3\Delta u-2s'((s-m_b^2)^2\pm s'(s'+2m_b^2-q^2))]\cr
&&-\lambda^2(2s'\pm q^2\mp m_b^2)+C_\pm \Big),
\end{eqnarray}
\begin{eqnarray}
\rho_g^{pert}=-{{3s'}\over{4\pi^2\lambda^{5/2}}}
\Big(\lambda(3m_b^2-2q^2-s)+3\Delta(3s'-s+q^2) \Big),
\end{eqnarray}
with
$u=s+s'-q^2,\lambda=u^2-4ss',\Delta=(s-m_b^2)(m_b^2-q^2)-s'm_b^2$,
$C_+=4s'(20s's\Delta+6\Delta\lambda-\lambda s's)$, and $C_-=0.$
(See Ref.\cite{yang} for the detailed calculation.) Note that in Eq. (25), the
contribution from the excited states has been approximately subtracted
due to the upper integral limit, $s_0^B$ or $s_0^\rho$, in the first term.
To calculate the contribution from the induced condensate, $\Pi^{induced}$,
we parametrize the bilocal condensate as in Ref.\cite{jung}:
\begin{eqnarray}
\langle\bar q(0){\cal G}(0,x) q(x)\rangle=\langle\bar q q \rangle g(x_E ^2),
\end{eqnarray}
with
\begin{eqnarray}
g(x^2_E)=\int_0^\infty d\alpha f(\alpha)e^{-x^2_E\alpha/4},
\end{eqnarray}
where the coordinates are Euclidean ($x^2_E=-x^2\geq 0$) and $g(x^2_E)$ is the
Euclidean space-time correlation function of the vacuum quarks.
Our choice of the vacuum function is
\begin{eqnarray}
f(\alpha)={4\over m_0^2}e^{-4\alpha/m_0^2},
\end{eqnarray}
where $m_0^2 \langle \bar qq\rangle$=
$-\langle g_s\bar q\sigma^{\mu\nu}G_{\mu\nu} q\rangle$. In this paper, we
have adopted the convention
$D^\alpha=\partial^\alpha+ ig_s {\lambda^a \over 2}A^\alpha_a$. If one adopts
$D^\alpha=\partial^\alpha- ig_s {\lambda^a \over 2}A^\alpha_a$, then
$m_0^2 \langle \bar qq\rangle$=
$\langle g_s\bar q\sigma^{\mu\nu}G_{\mu\nu} q\rangle$.
The corresponding correlation function $g(x^2)$ of the vacuum quarks
is of the monopole form
\begin{eqnarray}
g(x^2_E)={1\over 1+m_0^2 x^2_E/16 }.
\end{eqnarray}
Since $x^2$ has been assumed to be spacelike, we have $x^2_E=-x^2\geq 0$.
This choice leads to the empirical sea-quark distribution\cite{jung}.
By using the bilocal condensate parametrized above to Eq. (9), we obtain
the relevant induced condensate in our study as
\begin{eqnarray}
\langle b(0) {\cal G}(0,x) \gamma_\mu \gamma_5 u(x)\rangle_{\cal A}&=&
-i{\cal A}^\nu \Bigl(
\int d^4y e^{iqy}\langle 0|T(\bar b(0)\gamma_{\mu}\gamma_5 u(x),
\bar u(y)\gamma_\nu\gamma_5 b(y)) |0 \rangle\nonumber\\
&&\ \ \ - {\rm the\ perturbative\ term}\Big)\nonumber \\
&=&{\cal A}_\mu \int_0^1 e^{iqxu} \phi(u,x^2) du,
\end{eqnarray}
with
\begin{eqnarray}
\phi(u,x^2)=m_b \langle \bar uu \rangle \int_0^\infty d\beta{f(\beta)\over \beta}
e^{{-m_b^2\over \beta} ({1\over 1-u}-1)} e^{q^2 u/\beta} e^{{\beta\over 4}x^2(1-u)}.
\end{eqnarray}
In the following calculation, we adopt the approximation
$e^{{\beta\over 4}x^2(1-u)}\approx 1$ since $x^2\approx 0$.
Thus we obtain the contribution of the induced condensate,
as shown in Fig. 2(b),
\begin{eqnarray}
\Pi^{induced}_{A_1,Fig.2(b)}=\int_0^1 du {u\phi(u) \over (p'+uq)^2},
\end{eqnarray}
and $\Pi^{induced}_{A_2, Fig.2(b)}=\Pi^{induced}_{A, Fig.2(b)}=
\Pi^{induced}_{V, Fig.2(b)}=0$, where $\phi(u)$ =$\phi(u,x^2=0)$.
After performing the double Borel transform, Eq. (35) becomes
\begin{eqnarray}
{\bf B}[\Pi^{induced}_{A_1,Fig.2(b)}]=&&
{\bf B}\Bigl[\int_0^1 du {u\phi(u) \over u(p'+q)^2 +(1-u)p'^2+q^2 u(u-1)}\Bigr]
\nonumber\\
=&&-\phi\Bigl( {M'^2\over M^2+M'^2} ) \times {M'^4 M^2\over (M^2+M'^2)^2}
\times e^{q^2/(M^2+M'^2)}\nonumber
\end{eqnarray}
Note that the contribution of the induced
condensate in Eq. (35) was ignored in Ref. \cite{yang}.
The contribution of induced condensates as shown in Fig. 2(c) is given by
\begin{eqnarray}
\Pi^{induced}_{A_1, Fig.2(c)}(p,p';q^2)
=&&-{2g_s^2\langle\bar dd\rangle\over 9}
\chi^{\cal A}_{bu}(q^2){p^2-p'^2-q^2\over (p^2-m_b^2)p'^4},\nonumber\\
\Pi^{induced}_{A_2, Fig.2(c)}(p,p';q^2)
=&&{2g_s^2\langle\bar dd\rangle\over 9}
\chi^{\cal A}_{bu}(q^2){1\over (p^2-m_b^2)p'^4},\nonumber\\
\Pi^{induced}_{A, Fig.2(c)}(p,p';q^2)
=&&-{2g_s^2\langle\bar dd\rangle\over 9}
\chi^{\cal A}_{bu}(q^2){1\over (p^2-m_b^2)p'^4},\nonumber\\
\Pi^{induced}_{V, Fig.2(c)}(p,p';q^2)
=&&{4g_s^2\langle\bar dd\rangle\over 9}
\chi^{\cal V}_{bu}(q^2){1\over (p^2-m_b^2)p'^4},\nonumber\\
\end{eqnarray}
where we have approximated the non-local induced condensate as a local one and
the definition of $\chi^{\cal V(A)}$ is
\begin{eqnarray}
&&\langle\bar b(0)\sigma_{\alpha\beta} u(0)\rangle_{\cal V}\nonumber\\
&&=-i{\cal V}^\nu\Bigl( \int d^4y e^{iqy}\langle 0|T(\bar b(0)\sigma_{\alpha\beta} u(0),
\bar u(y)\gamma_\nu b(y)) |0 \rangle - {\rm the\ perturbative\ term}\Bigr) \nonumber\\
&&=-i{\cal V}^\nu  \chi^{\cal V}_{bu}(q^2)
(g_{\nu\alpha}q_\beta-g_{\beta\nu}q_\alpha),\nonumber\\
&&\langle\bar b(0)\sigma_{\alpha\beta}\gamma_5 u(0)\rangle_{\cal A}\nonumber\\
&&=-i{\cal A}^\nu\Bigl( \int d^4y e^{iqy}\langle 0|T(\bar b(0)\sigma_{\alpha\beta}
\gamma_5 u(0),
\bar u(y)\gamma_\nu\gamma_5 b(y)) |0 \rangle - {\rm the\ perturbative\ term}\Bigr) \nonumber\\
&&=-i{\cal A}^\nu \chi^{\cal A}_{bu}(q^2)
(g_{\nu\alpha}q_\beta-g_{\beta\nu}q_\alpha),
\end{eqnarray}
Similar to the case of two-point sum rules, we can write down $\chi^{\cal V(A)}$
as a dispersion relation form,
\begin{eqnarray}
\chi^{\cal V(A)}(q^2)=\int_0^\infty \frac{ds}{s-q^2}({\rm Im}\ \pi(s)-
{\rm Im}\ \pi^{pert}(s))\nonumber,
\end{eqnarray}
and further adopt the simple model:
\begin{eqnarray}
{\rm Im}\ \pi(s)=f^{\cal V(A)}\delta(s-m_{f^{\cal V(A)}}^2)+
{\rm Im}\ \pi^{pert}(s){\rm \theta}(s-s_0^{\cal V(A)}).\nonumber
\end{eqnarray}
The final values of $\chi^{\cal V(A)}$ are
\begin{eqnarray}
\chi^{\cal V(A)}_{bu}={f^{\cal V(A)}\over m_{f^{\cal V(A)}}^2-q^2}
-{3\over 8\pi}\int_{m_b^2}^{s_0} {\rm d}s {m_b\over s-q^2}\bigl[
1-2({m_b^2\over s})+({m_b^2\over s})^2\bigr],
\end{eqnarray}
where $f^{\cal V}$=0.28 GeV$^3$, $f^{\cal A}$=0.67 GeV$^3$,
$m_{f^{\cal V}}$=5.33 GeV, $m_{f^{\cal A}}$=5.71 GeV,
$s_0^{\cal V}$=33 GeV$^2$, and $s_0^{\cal A}$=38.2 GeV$^2$.
The detailed calculation is shown in Ref.\cite{yang}.
Note that the uncertainties of our sum rule results are well controlled
since in the following numerical analysis we find that the contribution
of the induced condensate in Eq. (35) is about 10$\%$ to the $A_1$ sum rule,
and that the contribution of Eq. (36) to relevant sum rules is less than 4$\%$.
The rest of the results for $ d_i$ is collected in the Appendix. And $d_3$,
$d_5$, $d_6^{(1)}$, $d_6^{(2)}$, and $d_6^{(3)}$ are represented pictorially
by Figs. 3(d), 3(e), 3(f), 3(g), and 3(h), respectively.

After equating the hadronic side to the quark side and then performing the
double Borel transform ${\bf B}[f]$ on $f$ in both variables $p^2$ and
$p'^2$, we derive the relevant form factor QCD sum rules as follows:
\begin{eqnarray}
{f_B m_\rho^2\over g_\rho} (m_B + m_\rho)
   e^{-(m_B^2/M^2)} e^{-(m_\rho^2/M'^2)}A_1(q^2)=&&
   {\bf B}[\Pi_{A_1}^{quark}],\\
{f_B m_\rho^2\over g_\rho}{1\over m_B + m_\rho}
   e^{-(m_B^2/M^2)} e^{-(m_\rho^2/M'^2)} A_2(q^2)=&&-
   {\bf B}[\Pi_{A_2}^{quark}],\\
{f_B m_\rho^2\over g_\rho} 2m_\rho^2
   e^{-(m_B^2/M^2)} e^{-(m_\rho^2/M'^2)}A(q^2)=&&
   {\bf B}[\Pi_{A}^{quark}],\\
{f_B m_\rho^2\over g_\rho}{2\over m_B + m_\rho}
   e^{-(m_B^2/M^2)} e^{-(m_\rho^2/M'^2)}V(q^2)=&&
   {\bf B}[\Pi_{V}^{quark}].
\end{eqnarray}

\section{RESULTS AND DISCUSSION}
\subsection{The determination of Borel windows}

In numerical analyses of form factors, people usually assume,
according to the empirical observation \cite{ioffe}, that
the suitable region of sum rules is at values of Borel masses twice
as large as that in the corresponding two-point functions\cite{bbd,neubert} and thus extract results
from larger Borel mass regions. However, we will provide a detailed
analysis to show that such choice is not always correct.

To obtain reliable Borel windows in which the form factor sum rules may be
safely used in the analysis, we apply the differential
operator $-M'^4 \partial\ \ell n/\partial M'^2$ to both sides of Eq. (39)
and then obtain the $\rho$ mass sum rule, which is free of the
parameters $g_\rho$, $f_B$, and the form factors ($A_1$, $A_2$, $A$, or $V$).
This procedure is usually used in analyzing the mass sum rules
in the two-point Green's function approach\cite{svz,yang2}.
Analogously, we also obtain the $\rho$ meson mass rules from Eqs. (40), (41),
 and (42). On the other hand, when applying the differential
operator $-M^4 \partial\ \ell n/\partial M^2$ to both sides of Eqs. (39-42),
we obtain four B meson mass sum rules. Totally, we obtain eight mass rules
(four for the $\rho$ meson and four for the B meson).
In the numerical analysis, we use the the following set of parameters:
\begin{eqnarray}
m_b=4.7\ {\rm GeV},\ \ m_u=m_d=0,\ \
\langle\bar uu\rangle=\langle\bar dd\rangle=-(240\ {\rm MeV})^3,\ \
m_0^2=0.6-0.8\ {\rm GeV}^2,
\end{eqnarray}
where $m_b$ is a pole mass. The heavy quark condensate will be dismissed
through this paper. To further improve the validity of the derived QCD
sum rules, we have performed the following replacements:
\begin{eqnarray}
\ m_0^2\langle\bar qq\rangle\to  m_0^2\langle\bar
qq\rangle L^{-{2\over{3b}}},\ \langle\bar qq\rangle\to\langle\bar qq\rangle
L^{4\over b},
\end{eqnarray}
with
\begin{eqnarray}
L={\hbox{\rm ln}((M'^2M^2)^{1/2}/\Lambda^2)\over \hbox{\rm
ln}(\mu^2/\Lambda^2)},\nonumber\\
\end{eqnarray}
$\Lambda$=100 MeV, $\mu$=500 MeV, and $b=11-2n_f/3$ with $n_f$ being the number
of $unfrozen$ quark flavors.
Note also that region $R^{-1}<<Q^2<<m_b^2$, where $R$ is
the confinement radius, exists an anomalous dimension $2/b$ for both
of the current operators\cite{vs,politzer,isgur2}, $\bar b\gamma_\nu\gamma_5q$
and $\bar b\gamma_\nu q$. In our numerical analysis of $B$ meson decay
form factors, the sum rules are
studied at $Q^2\sim\sqrt{M'^2M^2}\approx$ 3.4 GeV$^2$ [see below, Eq.(47)].
Therefore, to obtain physical form factors, which are independent of the
subtraction scale, we put the factor ${[\ell n(m_b^2/\Lambda^2)/
\ell n((M'^2M^2)^{1\over2}/\Lambda^2)]}^{12\over25}$ in the relevant form
factor sum rules (the right hand side of Eqs. (39-42)).
The requirement of obtaining reasonable mass results from these eight mass
sum rules puts severe constraints on the choices of the parameters,
$s_0^B$ and $s_0^\rho$. With a careful study, we find that the best
choice of parameters in our analysis is
\begin{eqnarray}
s_0^B=33-35\ {\rm GeV}^2,\ \ \ \ \ s_0^\rho=1.3-1.4\ {\rm GeV}^2.
\end{eqnarray}
Consequently we find that
\begin{eqnarray}
8.5\ {\rm GeV}^2 < M^2<11.5\ {\rm GeV}^2,\ \ \ \ \
0.9\ {\rm GeV}^2 < M'^2 < 1.2\ {\rm GeV^2}
\end{eqnarray}
are the reasonable choices through this paper.
The resultant masses are $m_\rho$=0.787$\pm$0.065 GeV and
$m_B$=5.11$\pm$0.18 GeV. The detailed results are collected in Table I.
As examples in Fig. 3(a) and Fig. 3(b), we plot the $\rho$ mass and B mass
(extracted from Eq.(39)), respectively, as a function of $M^2$ and $M'^2$.
Obviously, the study of both of the $\rho$ and B mass sum rules is a
gauge to understand the reliability of performing further numerical
analyses on the form factors.
 In concluding this subsection, we will to discuss two
``traditional" sum rule calculations by considering a three-point correlation
function in the existing literature:

The first is the work in Ref.\cite{bbd} in which the authors use the
vector current $\bar d \gamma^\nu u$  and
the pseudoscalar current $\bar b \gamma_5 d$ as the interpolating fields for
the $\rho$ meson and for the $B$ meson, respectively. For the form factor
$A_1$, they obtain the following sum rule:
\begin{eqnarray}
A_1 &&{f_B (m_B+m_\rho) m_B^2 m_\rho^2 \over m_b g_\rho} e^{-(m_B^2/M^2)}
e^{-(m_\rho^2/M'^2)} \nonumber\\
=&& {3\over 4\pi^2}\int_0 ^{s^B_0} ds\int_0 ^{s^\rho_0} ds'
 e^{-(s/M^2)} e^{-(s'/M'^2)}\Big ({m_b s'+m_u(s-m_b^2)\over 2\lambda^{1/2}}
\nonumber\\
&&\ \ +{m_b s'[(s-m_b^2)(q^2-m_b^2)+s' m_b^2]\over\lambda^{3/2}} \Bigr )\nonumber\\
-&&\langle \bar qq\rangle \Bigl (m_b m_u +{m^2-q^2\over 2}\Bigr )
e^{-m_b^2/M^2}\nonumber\\
+&& \langle \bar qq\rangle m_0^2\Bigl ({-1\over 6}+{(2m_b^2+3m_b m_u-2q^2)
(m_b^2-q^2)\over 12 M^2 M'^2}\nonumber\\
&&\ \ -{2m_b^2+3m_b m_u-2q^2\over 12 M'^2}
-{3m_b^2+9m_b m_u-4q^2\over 12 M^2}+m_b^2{m_b^2+2m_b m_u-q^2\over 8 M^4}
\Big ) e^{-m_b^2/M^2}\nonumber\\
+&&{16\alpha_s\pi \langle \bar qq\rangle ^2 m_b\over 9}
\Bigl ({8\over 9M'^2} -{1\over 9 M^2 } +{1\over (m_b^2-q^2)}
+{3m_b^2-2q^2\over 72 M^4}
-{m_b^2-q^2\over 6 M^2 M'^2}\nonumber\\
&&\qquad\qquad\qquad\qquad +{m_b^2(m_b^2-q^2)\over 72 M^6}
+{(m_b^2-q^2)^2\over 36 M^4 M'^2}
\Bigr) e^{-m_b^2/M^2},
\end{eqnarray}
where $\lambda$ is defined as before.
From their $A_1$ sum rule, one can extract, following the procedure shown
as above, the $\rho$ meson mass sum rule. The result is illustrated in
Fig. 4(a), from which we see that their result of $\rho$ mass sum
rule is not so stable as to determine the suitable Borel windows.
Moreover, one can easily find that once a Borel window is determined,
the resultant $\rho$ mass from their $A_2$ or $V$ sum rule is twice as
large as that from the $A_1$ sum rule. Note that if we adopt the
varying external field to do the calculation again, we find that
various induced condensates may enter complicatedly the sum rules
for $A_1$, $A_2$, and $V$ and the contribution from induced condensates
becomes a little big. Thus the estimate of the sum rule is less reliable.

The second calculation is done by Ball and Braun\cite{ball}. They use the tensor
current $\bar d \sigma^{\nu\sigma}u$ and
the pseudoscalar current $\bar b \gamma_5 d$ as the interpolating fields for
the $\rho$ meson and for the $B$ meson, respectively. For the form factor $A_2$,
they have
\begin{eqnarray}
A_2 &&{f_B f_\rho^\perp m_B^2\over m_b(m_B+m_\rho)} e^{-(m_B^2/M^2)}
e^{-(m_\rho^2/M'^2)} \nonumber\\
=&& {3\over 4\pi^2}\int_0 ^{s^B_0} ds\int_0 ^{s^\rho_0} ds'
 e^{-(s/M^2)} e^{-(s'/M'^2)} \Big ( {(s-m_b^2)u-2ss'\over \lambda^{3/2}}
\nonumber\\
&&\ \ -{(s-m_b^2)(s-m_b^2+2s')u^2
-3s'[(s-m_b^2)^2+2(s-m_b^2)s+ss']u \over \lambda^{5/2}}\nonumber\\
&&\ \  -{2ss'[(s-m_b^2)^2+2(s-m_b^2)s'+3ss']\over \lambda^{5/2}}\Bigr)\nonumber\\
-&& {m_b m_0^2\langle \bar qq\rangle \over 6 M^2 M'^2}e^{-m_b^2/M^2}\nonumber\\
-&& {16\alpha_s\pi \langle \bar qq\rangle ^2 \over 9}
\Bigl ({1\over 6M'^4} +{1\over 3 M^2 M'^2} +{m_b^2\over 36 M^4 M'^2}
      -{m_b^2-q^2\over 18 M^2 M'^4}\Bigr)e^{-m_b^2/M^2},
\end{eqnarray}
where $u$ and $\lambda$  are defined as before.
Note that, for this $A_2$ sum rules, one obtains the same result if
adopting the varying external field approach.
In Fig. 4(b) we show the $\rho$ mass sum rule extracted from Eq. (49).
From Fig. 4(b) we obtain $m_\rho^2\le 0$.
The result seems to indicate that  the tensor current
$\bar d \sigma^{\nu\sigma}u$ is not a good interpolating field for
the $\rho$ meson as mentioned in Ref.\cite{gov}. In the same reference,
the authors have pointed out that
the tensor current cannot easily produce a stable $\rho$ mass sum rule
if the vacuum saturation hypothesis goes in the opposite direction:
\begin{eqnarray}
\langle \bar q\gamma_5\lambda^a q\bar q\gamma_5\lambda^a q\rangle
=-{4\over 9}(1+\beta_5)\langle \bar qq\rangle^2,
\end{eqnarray}
with $\beta_5\approx -2$. Moreover, the authors\cite{gov} have also shown that
the threshold of the excited states in their $\rho$ meson mass sum rule
turns out to be unphysically low, the sum rule is
dominated by the continuum contribution, and the power corrections are
rather large.

\subsection{Numerical analysis of form factors}
To study numerically the form factors, we adopt the set of parameters
as shown in Eqs. (40), (43), and\cite{yang}
\begin{eqnarray}
f_B=160\ {\rm MeV},\ \ \ \ \  g_\rho=3.84.
\end{eqnarray}
The working Borel windows, which have been determined previously (Eq. (47)),
are 8.5 GeV$^2 < M^2 <$ 11.5 GeV$^2$ and 0.9 GeV$^2<M'^2<$ 1.2 GeV$^2$.
In these Borel ranges, the form factors are dominated by the leading
perturbative bare loop; for the $A_1$ form factor, the absolute value of
the contribution of $\langle g_s \bar d\sigma Gd\rangle$ is less than
50$\%$ of the bare loop. Moreover, the contribution of induced condensate
amounts to about 14$\%$ to the $A_1$ sum rule, while less than 4$\%$ to
$A_2$, $A$, or $V$. Note that unlike light-cone sum rules\cite{ali,ball},
we cannot apply ``directly" the heavy quark limit, $m_b\to \infty$,
to these sum rules since in that limit the series of the operator product
expansion (OPE) may become unconvergent. But in the case of the $B$ or $D$ meson,
our results indicate that the series of the OPE is in good convergence.
In Fig. 3(a-d), we plot the $A_1$, $A_2$, $A$, and $V$ form factors at $q^2=0$,
respectively, as a function of $M^2$ and $M'^2$.
We thus obtain the results on the form factors at $q^2=0$:
\begin{eqnarray}
A_1(0)&&=0.12\pm0.01,\nonumber\\
A_2(0)&&=0.12\pm0.01,\nonumber\\
A(0)&&=0.015\pm0.02,\nonumber\\
V(0)&&=0.15\pm0.02,
\end{eqnarray}
where the error comes from the variation in the Borel mass, $s_0^B$,
$s_0^\rho$, or $m_0^2$.
The resultant values of form factors are half as large as
the light-cone sum rule results\cite{ali,ball} or lattice QCD
calculation\cite{lat}.
We now consider the $q^2$ dependence of the various
form factors. The variation of the form factors with $q^2$ is of great interest,
since it probes the effects of strong interactions on the decay.
As the property of discontinuity in Ref.\cite{bbd} is mentioned (see
also the discussion below Eq. (24)),
the sum rules work well in the region $(m_Q^2-q^2)^2/q^2 > s'_0$. Therefore,
we could obtain the $q^2$ dependence of the form factors over a wide range
of $q^2$ (from $q^2$=0 up about to 9 GeV$^2$).
The $q^2$ dependence of our form factors is given by
\begin{eqnarray}
F(q^2)=F(0)(1-q^2/m^2_{F})^{-n},
\end{eqnarray}
where $n=1$ for $A_1$, $n=2$ for $A_2$, $A$, and $V$, and the fitted pole
masses are $m_{A_1}$=5.45 GeV, $m_{A_2}$=6.14 GeV, $m_A$=5.98 GeV and
$m_V$=5.78 GeV, respectively. Here the results are
evaluated at the central values of the Borel mass ranges in Eq.(47).
We find that our $q^2$ dependence of the form factors is well consistent
with the pole model ansatz by K$\ddot {\rm o}$rner and Schuler\cite{korner} and
recent lattice results\cite{lat} as well.
In the following calculation, we will extrapolate our $q^2$ dependence
of form factors to all possible kinematic region.
The pole model ansatz may be a good approximation for the form factor behavior
since it is consistent with this sum rule calculation in the region: 0 GeV$^2 <
q^2<$ 9 GeV$^2$ and also in good accordance with the
QCD power counting rules\cite{brodsky} at large $-q^2$ (the hard rescattering
region).
Moreover, by neglecting the light meson mass, we roughly obtain from
Eq. (53) the
relation $F(q_m^2)/F(0)\sim m_b^n$, where $q_m^2=(m_b-m_{\rho})^2$.
Therefore, our results
agree with the prediction of heavy quark symmetry\cite{isgur3}
 in the kinematic region near zero recoil $(q^2\approx q_m^2)$,
\begin{eqnarray}
a_++a_- &&\sim m_M^{-3/2},\ \ \ \ \ \ \ \ \ \ \ -a_++a_-\sim m_M^{-1/2},\nonumber\\
g&&\sim m_M^{-1/2},\ \ \ \ \ \ \qquad\qquad\quad\ f\sim m_M^{1/2}.
\end{eqnarray}
In Fig. 6(a) we plot the lepton-pair invariant mass spectrum
$d\Gamma/ d q^2$ of the $B \to \rho\ \ell\ \bar\nu_\ell$ decay together with
$d\Gamma_L/ d q^2$, $d\Gamma_+/ d q^2$, and $d\Gamma_-/ d q^2$.
The solid curve is for $d\Gamma/ d q^2$, and the long-dashed curve is for
$d\Gamma_L/ d q^2$, the portion of the rate with a longitudinal polarized
$\rho$ in the final state, the short-dashed curve is for $d\Gamma_-/ d q^2$,
the portion of the rate with a helicity minus $\rho$ in the final state,
while the dotted curve is for $d\Gamma_+/ d q^2$,
the portion of the rate with a helicity positive $\rho$ in the final state.
Similarly, in Fig. 6(b) we plot the electron spectrum $d\Gamma/d E_e$ of the
$B \to \rho\ \ell\ \bar\nu_\ell$ together with $d\Gamma_L/d E_e$, $d\Gamma_+/d E_e$,
and $d\Gamma_-/d E_e$. The solid curve is for $d\Gamma/d E_e$,
the long-dashed curve for $d\Gamma_L/d E_e$,
the short-dashed curve for $d\Gamma_-/d E_e$,
and the dotted curve for $d\Gamma_+/d E_e$.
Both of the Figs. 6(a) and 6(b) are plotted in the $B$ meson rest frame.

From Fig. 6 we obtain $d \Gamma/d q^2\approx d \Gamma_L/d q^2$ and
$d \Gamma/d E_e\approx d \Gamma_L/d E_e$ at the maximum recoil region
($q^2\approx 0$). This is quite reasonable because at low $q^2$, the
electron and antineutrino are nearly collinear, so that their net spin
along their motion is zero. Since the $B$ meson has spin zero, the energetic
recoiling $\rho$ meson must also have zero helicity.
The helicity minus contribution is more weighted towards the large $q^2$ value
than the helicity zero contribution around the small $q^2$ region.
Our results also show that $d \Gamma_-/d q^2\gg d \Gamma_+/d q^2$
as expected from the left-chiral $b_L\to u_L$ transition.
Our results for the decay rate are given by
\begin{eqnarray}
\Gamma(\bar B^0\to\rho^+ \ell^- \bar\nu_\ell)&&=
(5.1\pm1.0)\times|V_{ub}|^2\times 10^{12}  s^{-1},\cr
 \Gamma_L/\Gamma_T&&=0.85\pm0.03,\quad \Gamma_+/\Gamma_-=0.077\pm0.012,\cr
     \Gamma(\bar B^0\to\rho^+ \tau^- \bar\nu_\tau)&&=
(3.1\pm 0.06)\times|V_{ub}|^2\times 10^{12}  s^{-1}.
\end{eqnarray}
Since the induced condensate like Eq.~(35) does not contribute to the
sum rules in Ref.\cite{yang} for $\bar B^0\to \pi^+ \ell^- \bar\nu_\ell$,
therefore we take  $\Gamma(\bar B^0\to\pi^+ e^- \bar\nu_e)$=
$(5.4\pm1.6)\times|V_{ub}|^2\times 10^{12}  s^{-1}$
and  $\Gamma(\bar B^0\to\pi^+ \tau^- \bar\nu_\tau)$=
$(2.7\pm0.07)\times|V_{ub}|^2\times 10^{12}  s^{-1}$
from Ref.\cite{yang}.
We obtain the ratios $\Gamma(\bar B^0\to\rho^+ e^- \bar\nu_e)$/
$\Gamma(\bar B^0\to\pi^+ e^- \bar\nu_e)$$\approx$ 0.94 and
$\Gamma(\bar B^0\to\rho^+ \tau^- \bar\nu_\tau)$/
$\Gamma(\bar B^0\to\pi^+ \tau^- \bar\nu_\tau)$ $\approx$ 1.15.
Our result for the ratio
$\Gamma(\bar B^0\to\rho^+ e^- \bar\nu_e)$/ $\Gamma(\bar B^0\to\pi^+ e^- \bar\nu_e)$
is a bit smaller than, but still consistent with the
CLEO experimental value\cite{art,cleo} of 1.4$\pm$0.6.
However, one should note that in the CLEO experiment, the reconstruction
of the relevant events is model dependent. For instance, if the
ISGW II model is used, the ratio $\Gamma(\bar B^0\to\rho^+ e^- \bar\nu_e)$/
$\Gamma(\bar B^0\to\pi^+ e^- \bar\nu_e)$ is 1.1$\pm 0.7$.

In closing this subsection, we consider the case of the D meson decays.
By following the same procedure as the case of
$\bar B^0\to\rho^+\ell^-\bar\nu_\ell$, and using the parameters as in
Ref.\cite{yang}, we can obtain the same form factor results and their
$q^2$ behaviors as in Ref.\cite{yang}, except that $A_1^{D\to \rho}$~(0)
becomes 0.43$\pm$ 0.04 and $A_1^{D\to K^*}$(0) becomes 0.61$\pm$0.04.
The results for the form factors at $q^2=0$ are collected in Table II.
The calculated decay rates read
\begin{eqnarray}
\Gamma(D^0\to\rho^- \ell^+ \nu_\ell)&&=0.44\pm0.08 |V_{cd}|^2\times
10^{11} s^{-1}, \nonumber\\
 \Gamma_L/\Gamma_T&&=0.58\pm0.05,\quad \Gamma_+/\Gamma_-=0.041\pm0.002,
\nonumber\\
\Gamma(D^+\to K^{*0} \ell^+ \nu_\ell)&&=0.51\pm0.06|V_{cs}|^2\times
10^{11} s^{-1}, \nonumber\\
\Gamma_L/\Gamma_T&&=0.99\pm0.06,\quad \Gamma_+/\Gamma_-=0.19\pm0.02.
\end{eqnarray}
Taking $|V_{cs}|$=0.975, we obtain $B(D^+\to K^{*0} \ell^+ \nu_\ell)$
=5.1$\pm$0.7 $\%$. The experimental results\cite{pdg} are
$B(D^0\to K^{*-} \ell^+ \nu_\ell)$ =4.8$\pm$0.4 $\%$,
$\Gamma_L/\Gamma_T=1.23\pm0.13$, $\Gamma_+/\Gamma_-=0.16\pm0.04$.
Our results are consistent with the existing experimental data,
except that the value of $\Gamma_L/\Gamma_T$ is a little smaller than the
experimental data. Further applications of this approach to various exclusive
decay processes will be published elsewhere\cite{yang3}.

\section{SUMMARY}
In summary, we have used the varying external field approach of QCD
sum rules to compute the form factors for the
semileptonic decays $\bar B^0$ $\to \rho^+$ $\ell^-$ $\bar\nu_\ell$.
We have formulated this approach in a systematic way.
By extracting both of the $B$ meson and $\rho$ meson mass sum rules,
we can thus determine the reliable Borel windows in studying the relevant
form factor sum rules. We also include induced condensate contributions,
which have been ignored before, into the relevant sum rules.
Therefore, we demonstrate that some
QCD sum rule calculations in the literature are less reliable.
Our results strongly support the pole model ansatz by K$\ddot {\rm o}$rner
and Schuler on the $q^2$ dependence of the form factors. Combining with the
previous analysis in Ref.\cite{yang}, we obtain the ratio
$\Gamma(\bar B^0\to\rho^+ e^- \bar\nu_e)$/
$\Gamma(\bar B^0\to\pi^+ e^- \bar\nu_e)$$\approx$ 0.94 and
$\Gamma(\bar B^0\to\rho^+ \tau^- \bar\nu_\tau)$/
$\Gamma(\bar B^0\to\pi^+ \tau^- \bar\nu_\tau)$ $\approx$ 1.15.

\acknowledgments
The author would like to thank Prof. H.-Y. Cheng and Prof. Hoi-Lai Yu
for useful discussions and Prof. H.-Y. Cheng for a careful reading of
the manuscript. This work was supported in part by the National Science
Council of R.O.C. under Contract No. NSC86-2112-M01-020.

\appendix
\section{}
The quantities $d_3$, $d_5$, $d_6^{(1)}$, $d_6^{(2)}$, and $d_6^{(3)}$
defined in Eq. (25) read
\begin{eqnarray}
d_3^{A_1}=-{2m_u+m_d\over 2p'^2(p^2-m_b^2)}-{m_d\over 2 (p^2-m_b^2)^2}
 -{m_d(m_b^2-q^2)\over 2 p'^2(p^2-m_b^2)^2},
\end{eqnarray}

\begin{eqnarray}
d_5^{A_1}=&&{1\over 12}\Bigl[
-{6m_d m_b^2\over (p^2-m_b^4)^4}
-{2m_d \over (p^2-m_b^4)^3}
+{4m_b-2m_u+3m_d\over p'^2 (p^2-m_b^2)^2}
-{2m_u+m_d\over p'^4 (p^2-m_b^2)}\nonumber\\
&&-{2(m_d+m_u)(m_b^2-q^2)\over p'^4 (p^2-m_b^2)^2}
-{4(2m_b^2-q^2)m_d +6m_u m_b^2\over p'^2 (p^2-m_b^2)^3}
-{2m_d(m_b^2-q^2)^2\over p'^4 (p^2-m_b^2)^3}\nonumber\\
&&-{6m_d m_b^2 (m_b^2-q^2)\over p'^2 (p^2-m_b^2)^4}
\Bigr],
\end{eqnarray}

\begin{eqnarray}
d_6^{(1)A_1}=&&
+{1\over 81}\Bigl[
{3\over p'^4(p^2-m_b^2)}+{2\over p'^2(p^2-m_b^2)^2}-
{2\over (p^2-m_b^2)^3}\nonumber\\
&&-{10m_b^2+28m_bm_u-2q^2\over p'^2(p^2-m_b^2)^3}
 +{6(m_b^2-q^2)+7m_bm_u\over p'^4(p^2-m_b^2)^2}
+{6m_b^2\over (p^2-m_b^2)^4}\nonumber\\
&&+{2(m_b^2-q^2)(m_b^2+m_bm_u-q^2)\over p'^4(p^2-m_b^2)^3}
+{6m_b^2(m_b^2+m_bm_u-q^2)\over p'^2(p^2-m_b^2)^4}
\Bigr],
\end{eqnarray}

\begin{eqnarray}
d_6^{(2)A_1}=&&
{2\over 9}\Bigl[
{1\over p^4(p^2-m_b^2)}+{1\over p'^4 p^2}\nonumber\\
&&+{2m_um_b^2+m_b^2-q^2\over p^4 (p^2-m_b^2)p'^2}-
{4m_um_b^2\over p^2 (p^2-m_b^2)^2p'^2}
\Bigr],
\end{eqnarray}

\begin{eqnarray}
d_6^{(3)A_1}=&&
-{4\over 9} {1\over p'^4(p^2-m_b^2)},
\end{eqnarray}

\begin{eqnarray}
d_3^{A_2}={m_d\over 2p'^2 (p^2-m_b^2)^2},
\end{eqnarray}

\begin{eqnarray}
d_5^{A_2}=-{m_u \over 6p'^4(p^2-m_b^2)^2}
-{m_d \over 6p'^2(p^2-m_b^2)^3},
\end{eqnarray}
\begin{eqnarray}
d_6^{(1)A_2}=
-{1\over 27}\Bigl[
{2\over p'^2(p^2-m_b^2)^3}+{1\over p'^4(p^2-m_b^2)^2} \Bigr],
\end{eqnarray}
\begin{eqnarray}
d_6^{(2)A_2}=-{2\over 9m_b^4}\Bigl[
{1\over p'^2(p^2-m_b^2)}-{1\over p'^2 p^2}- {m_b^2\over p'^2 p^4}\Bigr],
\end{eqnarray}

\begin{eqnarray}
d_6^{(3)A_2}=0,
\end{eqnarray}

\begin{eqnarray}
d_3^{A}=-{m_d\over 2p'^2 (p^2-m_b^2)^2},
\end{eqnarray}
\begin{eqnarray}
d_5^{A}= {m_u \over 6p'^4(p^2-m_b^2)^2}
-{m_d \over 2p'^2(p^2-m_b^2)^3},
\end{eqnarray}

\begin{eqnarray}
d_6^{(1)A}=-{1\over 81}\Bigl[
 {2\over p'^2(p^2-m_b^2)^3}-{3\over p'^4(p^2-m_b^2)^2}
\Bigr],
\end{eqnarray}
\begin{eqnarray}
d_6^{(2)A}={2\over 9m_b^4}\Bigl[
{1\over p'^2(p^2-m_b^2)}-{1\over p'^2 p^2}-
{m_b^2\over p'^2 p^4}\Bigr],
\end{eqnarray}

\begin{eqnarray}
d_6^{(3)A}=0,
\end{eqnarray}

\begin{eqnarray}
d_3^{V}=-{m_d\over p'^2(p^2-m_b^2)^2},
\end{eqnarray}

\begin{eqnarray}
d_5^V= {m_u \over 3p'^4(p^2-m_b^2)^2}
-{m_d \over 3p'^2(p^2-m_b^2)^3}
-{m_d m_b^2\over 2p'^2(p^2-m_b^2)^4}
-{m_d (m_b^2-q^2) \over 6p'^4(p^2-m_b^2)^3},
\end{eqnarray}
\begin{eqnarray}
d_6^{(1)V}=&&{1\over 81}\Bigl[
{4\over p'^2(p^2-m_b^2)^3}-{1\over p'^4(p^2-m_b^2)^2}
- {6m_b^2\over p'^2(p^2-m_b^2)^4}-{2(m_b^2-q^2)\over p'^4(p^2-m_b^2)^3}\Bigr],
\end{eqnarray}
\begin{eqnarray}
d_6^{(2)V}=
{4\over 9m_b^4}\Bigl[
{1\over p'^2(p^2-m_b^2)}-{1\over p'^2 p^2}-
{m_b^2\over p'^2 p^4}\Bigr],
\end{eqnarray}

\begin{eqnarray}
d_6^{(3)V}=0.
\end{eqnarray}

%
%


\begin{table}
\begin{center}
\caption{Estimates of the $\rho$ and B meson masses from Eqs.(39-42).}
\begin{tabular}{ccc}
&{$m_\rho$ (GeV)}& $m_B$ (GeV)\\ \hline
Eq. (39) &0.722$\sim$ 0.806 &5.15 $\sim$ 5.28\\
Eq. (40) &0.820$\sim$ 0.844 &4.99 $\sim$ 5.02\\
Eq. (41) &0.830$\sim$ 0.852 &4.92 $\sim$ 4.95\\
Eq. (42) &0.764$\sim$ 0.803 &4.98 $\sim$ 5.01\\
Average&0.787$\pm$0.064&5.10$\pm$0.19\\
\end{tabular}
\end{center}

\begin{center}
\caption{Estimates of various weak semileptonic form factors
for $D\to \rho, K^*$.}
\begin{tabular}{ccccc}
 &$A_1(0)$ &$A_2(0)$ &$A(0)$ &$V(0)$\\ \hline
$D^0\to \rho^- $     &0.43$\pm$0.04 &0.57$\pm$0.08
                     &0.30$\pm$0.07 &0.98$\pm$0.11\\
$D^0\to K^{*-}$     &0.61$\pm$0.04 &0.67$\pm$0.08
                              &0.22$\pm$0.03&1.10$\pm$0.10\\
\end{tabular}
\end{center}
\end{table}

\newpage
\centerline{Figure Captions}
\noindent
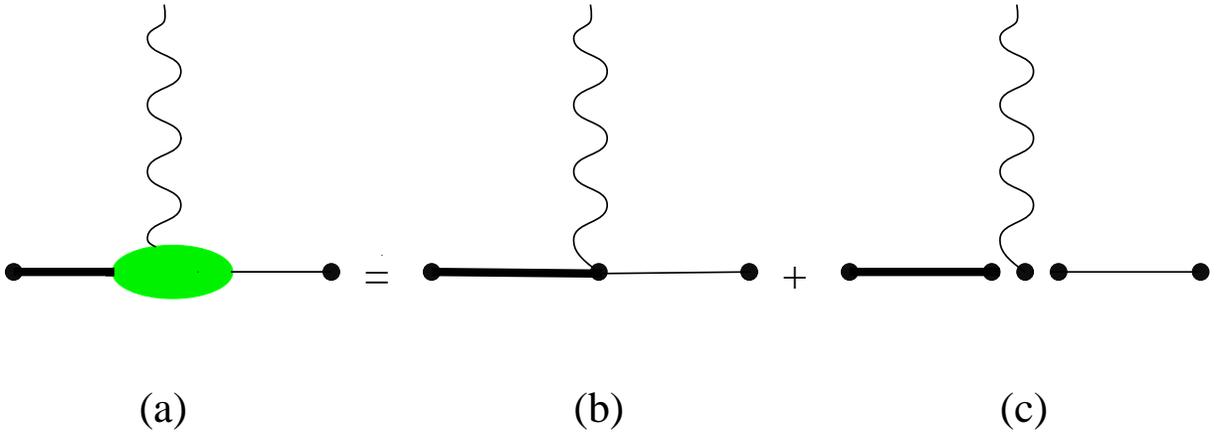
Fig. 1. Diagrams for the propagator of Eq. (4).
The heavy quark propagator is represented by the heavy line.
\vskip 2.5mm
\noindent
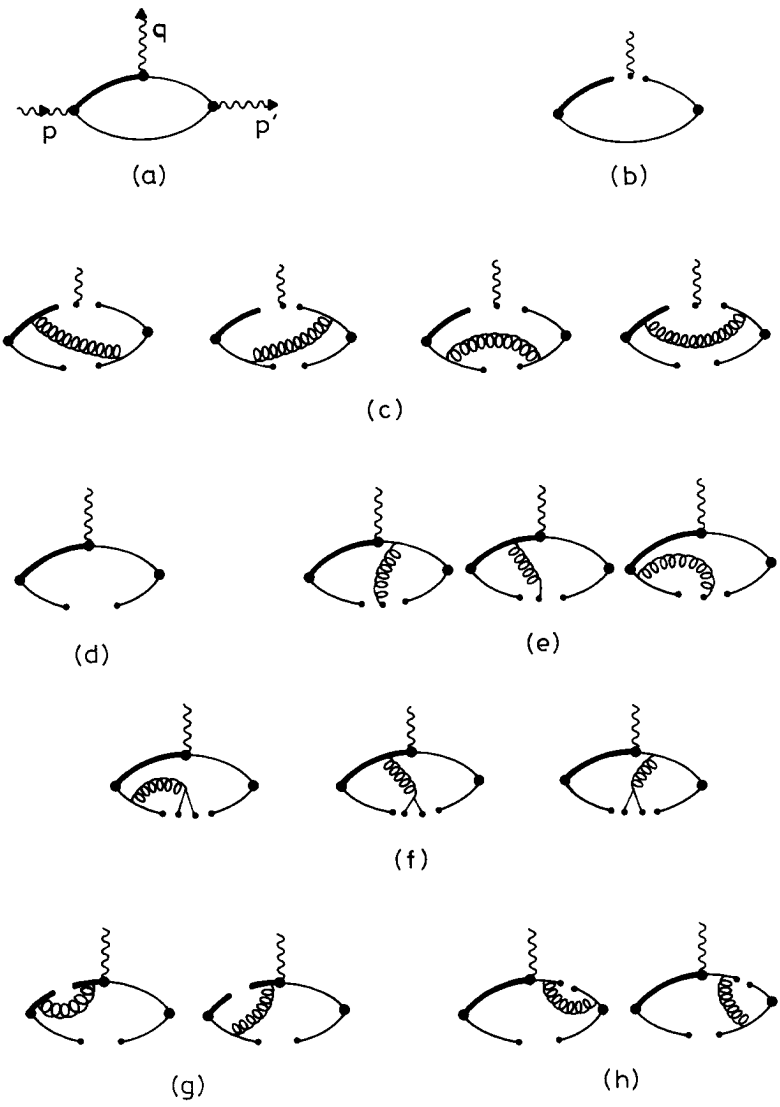
Fig. 2. Diagrams for the correlation function of Eq. (20).
The heavy quark propagator is represented by the heavy line.
\vskip 2.5mm
\noindent
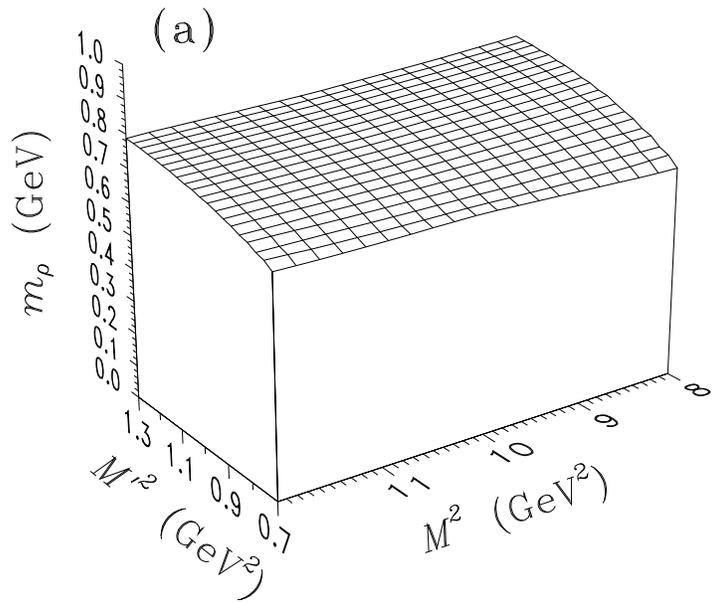
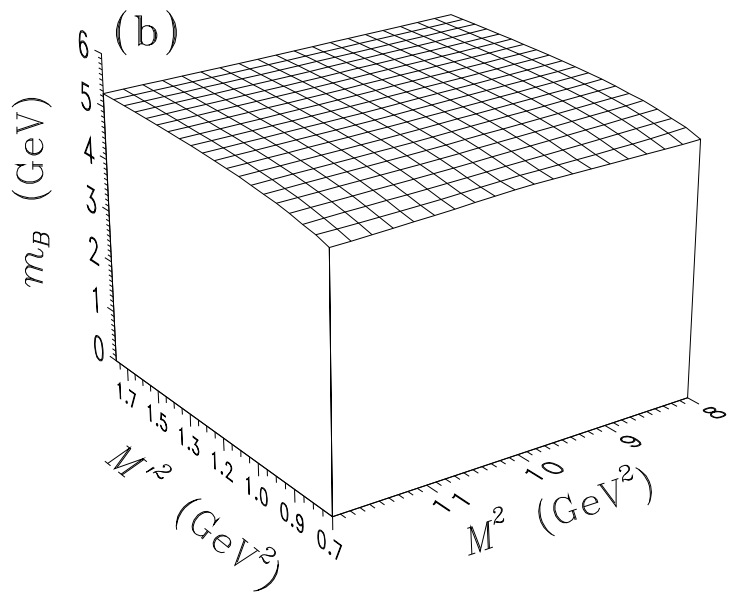
Fig. 3.
The $\rho$ mass and $B$ mass, extracted from Eq. (39), plotted
as a function of the square of the Borel masses $M^2$ and $M'^2$.

\vskip 2.5mm
\noindent
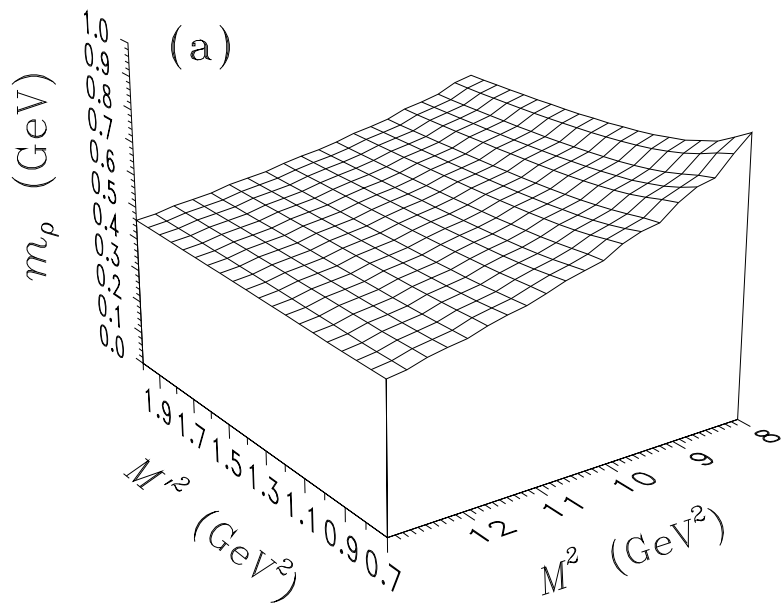
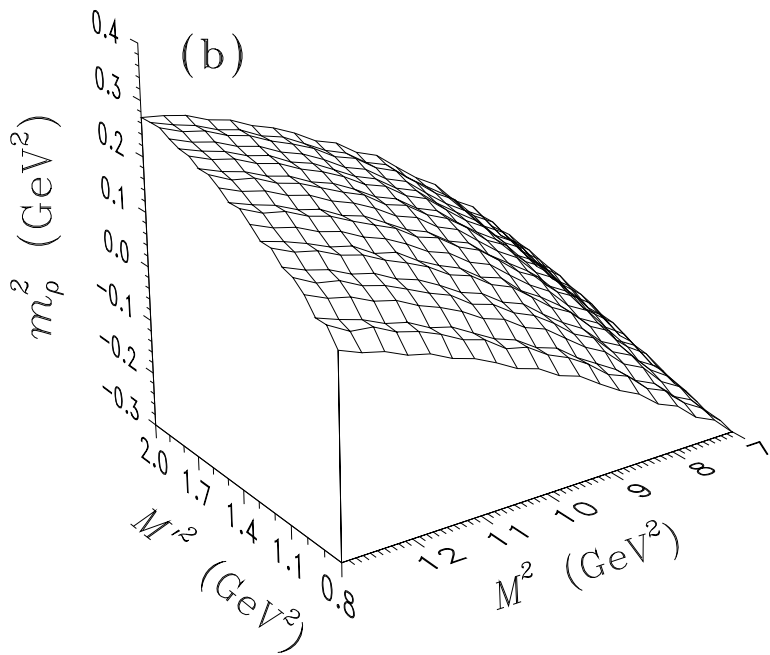
Fig. 4.\hskip 3mm
The $\rho$ mass as a function of the square of the Borel masses
$M^2$ and $M'^2$.
In Fig. 4(a) the $\rho$ mass is extracted from Eq. (48), the $A_1$ sum rule
of \cite{bbd}.
In Fig. 4(b) the $\rho$ mass is extracted from Eq. (49), the $A_2$ sum rule
of \cite{ball}.

\vskip 2.5mm
\noindent
Fig. 5.
The $A_1$(0), $A_2$(0), $A$(0), and $V$(0) form factors
plotted as a function of the square of the Borel masses $M^2$ and $M'^2$.

\vskip 2.5mm
\noindent
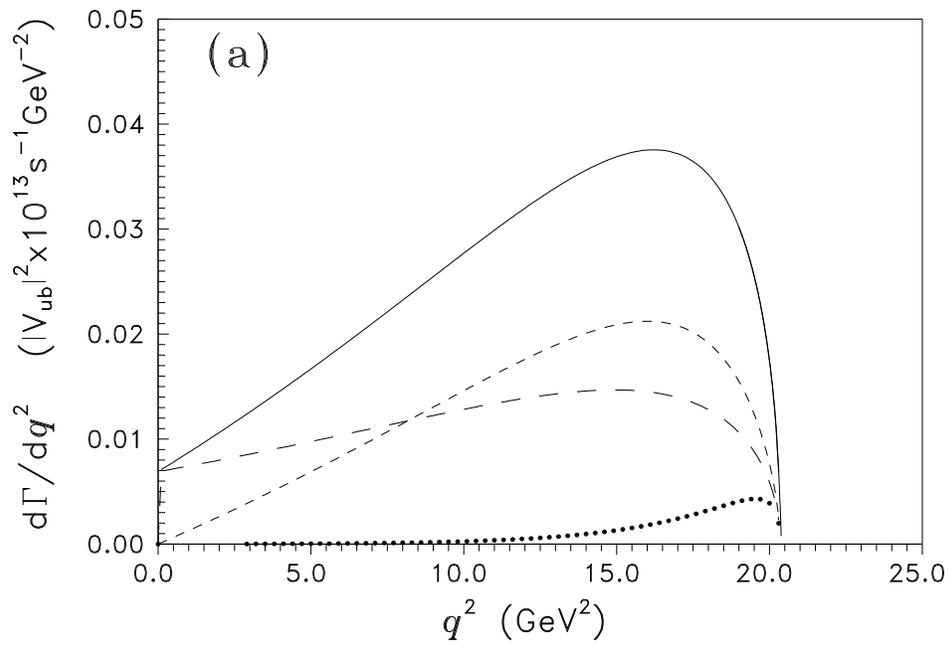
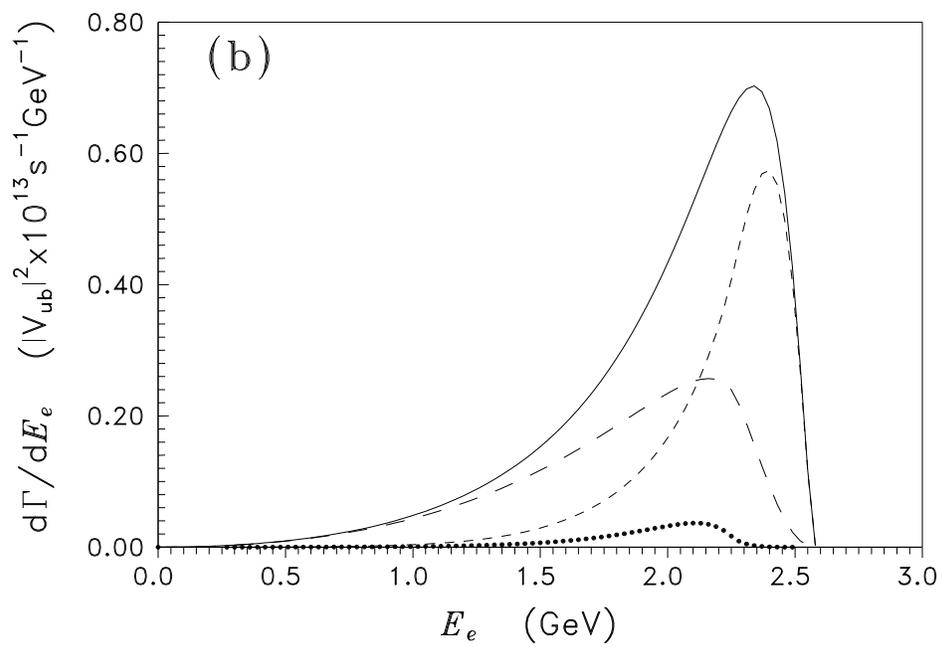
Fig. 6.
(a) The lepton-pair invariant mass spectra $d\Gamma/dq^2$ plotted as a
function of $q^2$.
 The solid curve is $d\Gamma/ d q^2$. The long-dashed curve stands for
$d\Gamma_L/ d q^2$, the portion of the rate with
a longitudinal polarized $\rho$ in the final state.
The short-dashed curve stands for
$d\Gamma_-/ d q^2$, the portion of the rate with
a helicity minus $\rho$ in the final state,
while the dotted curve is for $d\Gamma_+/ d q^2$,
the portion of the rate with a helicity positive $\rho$ in the final state.
\noindent
(b) The electron spectra $d\Gamma/d E_e$ of the
$\bar B \to \rho^+\ \ell^-\ \bar\nu_\ell$ together with $d\Gamma_L/d E_e$,
$d\Gamma_+/d E_e$, and $d\Gamma_-/d E_e$. The solid curve, the long-dashed
curve, the short-dashed curve, and the dotted curve are for
$d\Gamma/d E_e$, $d\Gamma_L/d E_e$, $d\Gamma_-/d E_e$, and
$d\Gamma_+/d E_e$, respectively.


\begin{thebibliography}{99}
\bibitem{bbd} P. Ball, V. M. Braun, and H. G. Dosch, Phys. Rev. D{\bf 44}, 3567
(1991); P. Ball, ibid, {\bf 48}, 3190 (1993).

\bibitem{bel} V. M. Belyaev, A. Khodjamiryan, and R. Ruckl,
Z. Phys. C{\bf 60}, 349 (1993).

\bibitem{ali} A. Ali, V. M. Braun, and H. Simma,
Z. Phys. C{\bf 63}, 437 (1993).

\bibitem{lat} J. M. Flynn et al. (UKQCD coll.), Nucl. Phys. B{\bf 461}, 327
(1996); ibid, B{\bf 476}, 313 (1996).

\bibitem{korner} J.G. K$\ddot {\rm o}$rner and G. A. Schuler, Z. Phys. C{\bf 46}, 93
(1990).

\bibitem{wsb} M. Wirbel, B. Stech, and M. Bauer, Z. Phys. C{\bf 29}, 637
(1985).

\bibitem{isgur1} N. Isgur, D. Scora, B. Grinstein, and M. Wise, Phys. Rev.
D{\bf 39}, 799 (1989).

\bibitem{scora} D. Scora and N. Isgur, Phys. Rev. D{\bf 52}, 2783 (1995).

\bibitem{cheng}
H.-Y. Cheng, C.-Y. Cheung, and C.-W. Hwang, Phys. Rev. D{\bf 55}, 1559 (1997).

\bibitem{yang} K.-C. Yang and W-Y. P. Hwang, Z. Phys. C{\bf 73}, 275 (1997).

\bibitem{ball} P. Ball and V. M. Braun, Phys. Rev. D{\bf 55}, 5561 (1997).

\bibitem{yu}
H.-n. Li and H.-L. Yu, Phys. Rev. Lett. {\bf 74}, 4388 (1995);
H.-n. Li and H.-L. Yu, Phys. Lett. B{\bf 353}, 301 (1995).

\bibitem{rad}
A. V. Radyushkin and R. Ruskov, Nucl. Phys. B{\bf 481}, 625 (1996).

\bibitem{smilga} A. V. Smilga and M. A. Shifman, Yad. Fiz. {\bf 37},
1613 (1983) [ Sov. J. Nucl. Phys. {\bf 37}, 958 (1983)].

\bibitem{yang2} K.-C. Yang, W-Y. P. Hwang, E. M. Henley,
and L. S. Kisslinger, Phys. Rev. D{\bf 47}, 3001 (1993).

\bibitem{cut} R. E. Cutkosky, J. Math. Phys. {\bf 1}, 429 (1960).

\bibitem{ex} One should note that not all tensor structures are independent. In
the case of deriving the form factor $g$, we choose $S_{\alpha\mu\nu}$,
$S'_{\alpha\mu\nu}$, $p_\mu T_{\nu\alpha}-p_\alpha T_{\mu\nu}$,
$p'_\mu T_{\nu\alpha}-p'_\alpha T_{\mu\nu}$, $p_\nu T_{\alpha\mu}$,
and $p'_\nu T_{\alpha\mu}$ as independent structures with
$T_{\alpha\mu}=\epsilon_{\alpha\mu\beta\rho}p'^\beta
p^\rho$, $S_{\alpha\mu\nu}=\epsilon_{\alpha\mu\nu\beta}p^\beta$, and
$S'_{\alpha\mu\nu}=\epsilon_{\alpha\mu\nu\beta}p'^\beta$. See also
Ref.\cite{yang}.

\bibitem{jung} H. Jung and L. S. Kisslinger, Nucl. Phys. A{\bf 586},
682 (1995). The analytic form of the non-local condensate is uncertain.
If a simple Gaussian-type mode for the coordinate dependence of the
non-local condensate is adopted $f(\alpha)$=$\delta(\alpha-m_0^2/4)$,
$\langle \bar q(0)q(x)\rangle$=$\langle \bar q(0)q(0)\rangle$$\times
e^{-x_E m_0^2/16}$
(S. V. Mikhailov and A. V. Radyushkin, Yad. Fiz. {\bf 49},794 (1989)
[Sov. J. Nucl. Phys. {\bf 49}, 494 (1989)]), the contribution of this
induced condensate to the $A_1$ from factor is almost zero.
However, Jung and Kisslinger have shown that if the simple Gaussian-type
mode for the coordinate dependence of the non-local condensate is
used, the resulting quark distribution is $xq(x)\propto \delta(x-8m_0^2/M^2)$,
which is not acceptable($M^2$:the Borel mass).
Moreover, since the induced condensates are not so important in my
sum rules, the resultant uncertainties are thus reasonably small.
\bibitem{ioffe}
B. L. Ioffe and A. V. Smilga, Phys. Lett. B{\bf 114}, 353 (1982);
B. L. Ioffe and A. V. Smilga, Nucl. Phys. B{\bf 216}, 373 (1983).

\bibitem{neubert} M. Neubert, Phys. Rev. D{\bf 45}, 2451 (1991).

\bibitem{svz} M. A. Shifman, A. I. Vainshtein, and V. I. Zakharov, Nucl. Phys.
B{\bf 147}, 385, 448 (1979); L. J. Reinders, H. Rubinstein, and S. Yazaki,
Phys. Rep. {\bf 127}, 1 (1985).

\bibitem{vs} M. B. Voloshin and M. A. Shifman, Yad. Fiz. {\bf 45}, 463
(1987)[Sov. J. Nucl. Phys. {\bf 45}, 292 (1987)].

\bibitem{politzer} H. D. Politzer and M. B. Wise, Phys. Lett. B{\bf 206}, 681
(1988); ibid, {\bf 208}, 504 (1988).

\bibitem{isgur2} N. Isgur and M. B. Wise, Phys. Lett. B{\bf 232}, 113 (1989);
{\bf 237}, 527 (1990).

\bibitem{gov} J. Govaerts, et al., Nucl. Phys. B{\bf 283}, 706 (1987).

\bibitem{brodsky}
S. J. Brodsky and G. P. Lepage, Phys. Rev. D{\bf 22}, 2157 (1980).

\bibitem{isgur3} N. Isgur and M. B. Wise, Phys. Rev. D{\bf 42}, 2388 (1990).

\bibitem{art} M. Artuso, HEPSY 96-2, hep-ex/9610009.

\bibitem{cleo} J. P. Alexander et al. (CLEO Collaboration),
Phys. Rev. Lett. {\bf 77}, 5000 (1996).

\bibitem{pdg} Particle Data Group, Phys. Rev. D{\bf 54}, 1 (1996).

\bibitem{yang3} K.-C. Yang, in preparation.
\end{thebibliography}
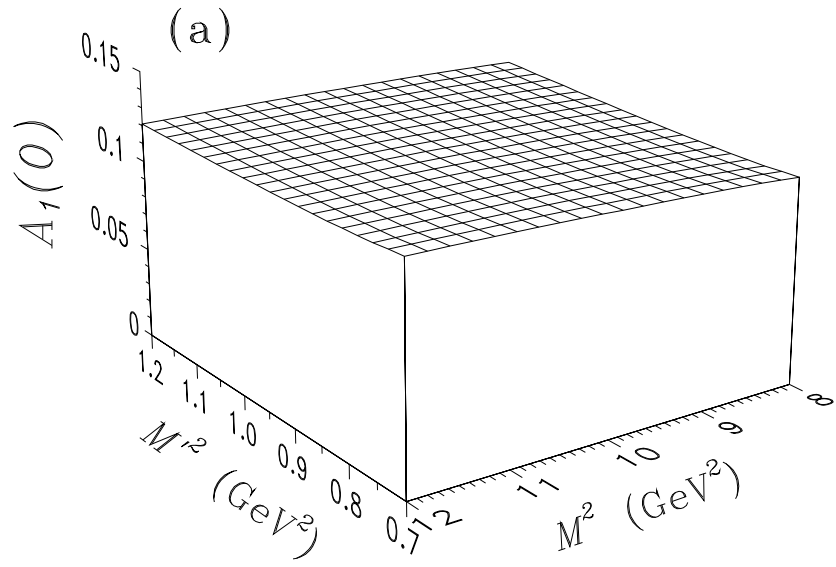
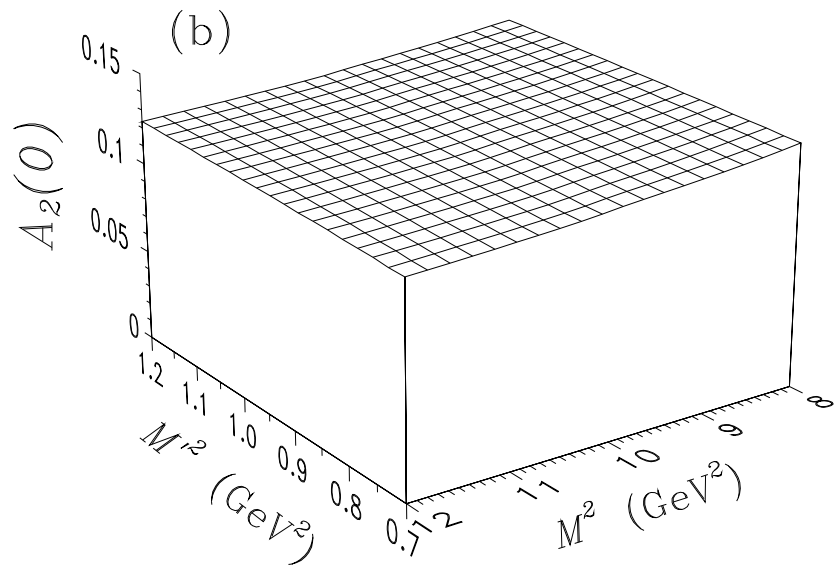
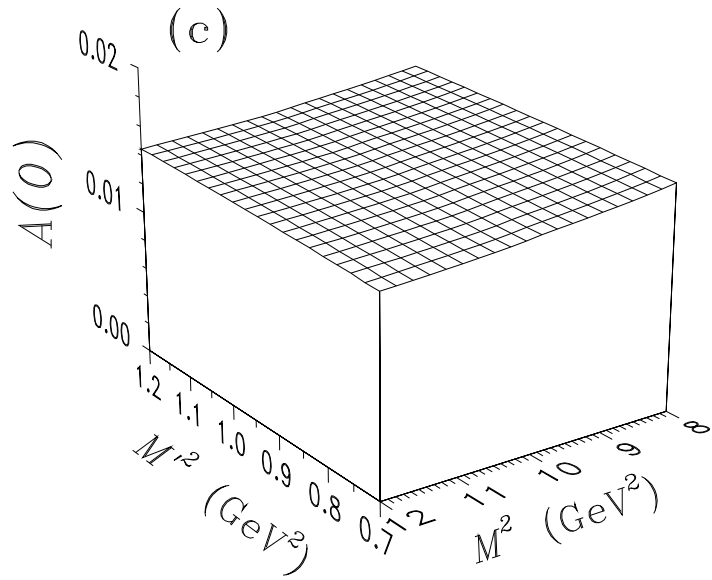
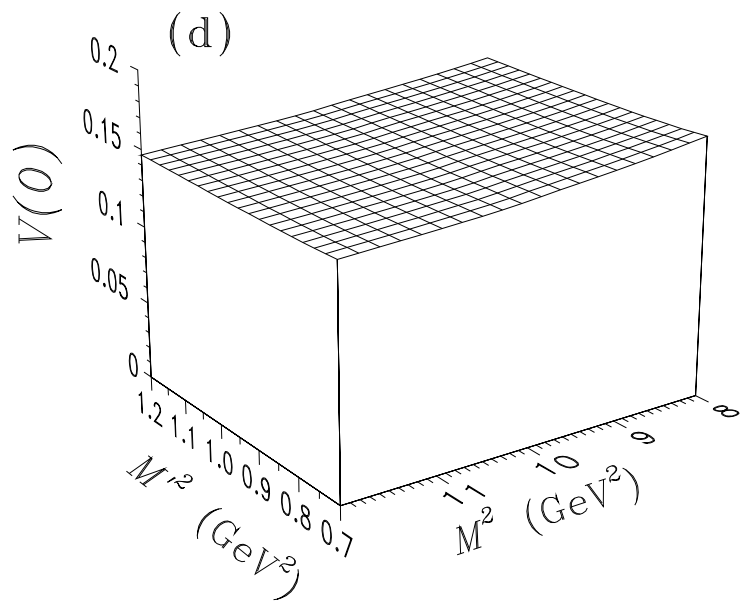
\end{document}